\documentclass[letterpaper,twocolumn,10pt]{article}
\usepackage{usenix2019_v3}
\usepackage{cite}
\usepackage{amsmath,amssymb,amsfonts, comment}
\usepackage{algorithmic,comment,todonotes}
\usepackage[linesnumbered,ruled,vlined]{algorithm2e}
\usepackage{graphicx}
\usepackage{url}

\usepackage{breakurl}
\usepackage{xspace}
\usepackage{titlesec}
\usepackage[title]{appendix}
\usepackage{caption}
\usepackage{subcaption}
\usepackage{xcolor}
\usepackage{soul}
\usepackage{tikz}
\usepackage{multirow}
\usepackage{multicol}
\usepackage[normalem]{ulem}
\usepackage{authblk}

\usepackage{textcomp}
\def\BibTeX{{\rm B\kern-.05em{\sc i\kern-.025em b}\kern-.08em
    T\kern-.1667em\lower.7ex\hbox{E}\kern-.125emX}}

\begin{document}
\newcommand{\newchange}{}
\newcommand{\newer}{}
\newcommand\older[1]{}
\newcommand{\says}[3]{\todo[size=\small,color=#2,inline]{#1 says: #3}}
\newcommand{\hide}[1]{{\sethlcolor{black}\hl{#1}}}

\newcommand\akash[1]{\says{Akash}{cyan}{#1}}
\newcommand\mani[1]{\says{Mani}{orange}{#1}}
\newcommand\luis[1]{\says{Luis}{pink}{#1}}
\newcommand\joseph[1]{\says{Joseph}{yellow}{#1}}

\newcommand{\orange}[1]{\textcolor{orange}{#1}}
\newcommand{\green}[1]{\textcolor{green}{#1}}
\newcommand{\blue}[1]{\textcolor{blue}{#1}}
\newcommand{\red}[1]{\textcolor{red}{#1}}


\newcommand{\name}{\textsc{SnoopDog\hspace{3pt}}}
\newcommand{\namenospace}{\textsc{SnoopDog}\xspace}
\newcommand{\nametitle}{SnoopDog}

\newcommand{\etal}{~\emph{et al.}\xspace}
\newcommand{\wifi}{Wi-Fi\xspace} 

\newcommand*\circled[1]{\tikz[baseline=(char.base)]{
            \node[shape=circle,fill,inner sep=1pt] (char) {\textcolor{white}{#1}};}}

\title{I Always Feel Like Somebody's Sensing Me! \\A Framework to Detect, Identify, and Localize Clandestine Wireless Sensors}

\author[ ]{Akash Deep Singh$^\dagger$}
\author[ ]{Luis Garcia$^{\dagger,\S}$}
\author[ ]{Joseph Noor$^\dagger$}
\author[ ]{Mani Srivastava$^\dagger$}
\affil[ ]{\textit {akashdeepsingh@g.ucla.edu, lgarcia@isi.edu, jnoor@cs.ucla.edu,} and \textit{mbs@ucla.edu}}
\affil[ ]{$^\dagger$University of California, Los Angeles (UCLA), $^\S$ USC ISI (work done at $^\dagger$)}

\maketitle

\begin{abstract}
The increasing ubiquity of low-cost wireless sensors has enabled users to easily deploy systems to remotely monitor and control their environments. However, this raises privacy concerns for third-party occupants, such as a hotel room guest who may be unaware of deployed clandestine sensors. Previous methods focused on specific modalities such as detecting cameras, but do not provide a generalized and comprehensive method to capture arbitrary sensors which may be ``spying" on a user. In this work, \newchange{we propose \namenospace, a framework to not only detect common \wifi based wireless sensors} that are actively monitoring a user, but also classify and localize each device. 
\name works by establishing causality between patterns in observable wireless traffic and a trusted sensor in the same space, e.g., an inertial measurement unit (IMU) that captures a user's movement. Once causality is  established, \name performs packet inspection to inform the user about the monitoring device. Finally, \name localizes the clandestine device in a 2D plane using a novel trial-based localization technique. We evaluated \name across several devices and various modalities, and were able to detect causality for snooping devices $95.2\%$ of the time, and localize devices to a sufficiently reduced sub-space.
\end{abstract}



\section{Introduction}\label{sec:intro}

The proliferation of low-cost wireless sensors has facilitated increased adoption into smart home, building, and city deployments\cite{staff_2018, heater_2019}.
Although there are profound positive impacts that ubiquitous sensor-rich environments can have on society, there is an inherent risk in enabling users access to such pervasive sensing, particularly when these environments host occupants oblivious to the presence of these sensors.


An individual's privacy in these contexts is entirely at the discretion of the owner. Regulation is unclear in informal settings, such as a guest residing in a homestay lodging. 
There have been reported instances where a hosting owner has attempted to spy on homestay occupants~\cite{airbnb-news}, motel lodgings~\cite{jeong_griffiths_2019}, and rooms aboard cruise ships \cite{staff_2018_1}. There are even instances in well-established hotel chains and mall restrooms when a malicious employee or customer has bugged several rooms~\cite{virginia_2019}. \newchange{Beyond commercial applications}, Southworth\etal report that domestic abusers may use such sensors for intimate partner stalking\cite{southworth2007intimate}. \newchange{Thus, potential victims with privacy concerns must take a proactive approach to detect clandestine sensors.}

The prevalent method to detect bugs relies on an RF receiver that senses if the received power in a particular frequency range is above a certain threshold. However, as bug detectors work on the principle of sensing surrounding RF signals, they can easily be triggered by legitimate RF devices such as mobile phones, radios, smart TVs, and other smart devices, thus limiting the practicality of these detectors. \newchange{An alternate method has emerged to detect the presence of IoT devices based on network traffic statistics~\cite{huang2019iot}. However, these methods only ascertain the presence of a device without semantic information regarding device information, location, or whether the device is actually monitoring a user.}  

More sophisticated solutions have since emerged targeting wireless cameras specifically.
Wampler\etal~\cite{wampler2015information} showed that changing lighting conditions causes notable variations to appear in a wireless camera's video traffic; that is, video encoding leaks sensitive environmental information. 
Flickering a light source for a short period of time can then be used in correlation with network traffic changes to identify hidden cameras~\cite{nassi2019drones, liu2018detecting}. Similarly, an approach has been presented that correlates the \wifi traffic patterns of a trusted camera with \wifi traffic patterns of other hidden cameras on a network to detect whether they are simultaneously observing the same space~\cite{wu2019you}. Unfortunately, these camera-specific approaches fail to generalize across modalities.
For example, varying lighting conditions would be ineffective for detecting a hidden microphone or an RF sensor. In recent work, human motion was used to detect a hidden camera with coarse localization (i.e., indoors or outdoors)~\cite{cheng2018dewicam}. We argue that human motion is an emblematic event to generalize across modalities, as the objective in revealing bugs is typically to determine if the user is being observed. 



In this paper, we propose \namenospace, a generalized framework to detect clandestine wireless sensors monitoring a user in a private space. \name leverages the notion of causality to determine if the values of a trusted sensor cause patterns in \wifi traffic stemming from other devices. 
In particular, \name works by having the user perturb the trusted sensor values to observe if there is a causal pattern in the \wifi traffic for a different device. For instance, if a wireless camera \newchange{or a motion detector} is monitoring a user who is wearing an inertial measurement unit (IMU), the IMU values will indicate a causal relationship with the camera's \wifi traffic. \name utilizes encoding scheme models of different wireless sensing modalities to classify the sensor type, and then cross-references packet headers with publicly available information of manufacturers to identify the specific device model. We further introduce a novel fine-grained localization approach that leverages sensor coverage techniques to locate a detected sensor. We implemented \name using a user's mobile phone for ground truth sensors and a laptop for sniffing \wifi traffic patterns. In the future, we envision \name to be implemented entirely as an app on either a smartwatch or a smartphone, both of which have sufficient sensing capabilities, but currently require \wifi card improvements to allow for channel hopping in monitor mode, thus making \name easily accessible to non-technical users.


\name operates in two stages. 
\name begins in a \textit{passive} monitoring phase that searches for suspicious causal patterns between the wireless traffic and the user's normal activity with their smartphone or wearable device.
If a device is flagged as potentially monitoring the user, an \textit{active} phase is engaged, and the user is instructed to perform a series of specific actions to detect the sensor with high accuracy. During the active phase, localization can optionally be engaged to find the clandestine sensor. The user can either skip the background or the active phase as per their convenience.

We evaluate \name over a representative set of wireless sensors following a taxonomy of popular sensing devices that may be used for surveillance. 
The framework had a detection rate of $96.6\%$ and a device classification rate of $100\%$ when the injected multi-modal event was human motion.  
We show that the location of the bug can be narrowed down to a sufficiently reduced region that easily facilitates a user's search. 
This feature is a notable improvement over existing approaches that only localize devices as either indoors or outdoors. 
While \name cannot detect \textit{any} wireless sensor monitoring the user (Section \ref{sec:limitations}), it can detect a broad set of commonly used wireless sensors~\cite{amazoncameras,amazonhomeautomation, ding2011sensor}. 





\noindent\textbf{Contributions:} Our contributions are summarized as follows:
\begin{itemize}
  \item We propose \namenospace, the first generalized framework to detect hidden clandestine sensors, including video, audio, motion, and RF. \name leverages the cause-effect relationship between a trusted set of sensor values and \wifi traffic patterns when observing a multi-modal injected event.
  \item We present a novel technique that leverages the notion of directional sensor coverage to provide state-of-the-art localization for clandestine devices.
  \item We show how \name can reveal device information by cross-referencing packet inspection with publicly available device manufacturer information.
  \item We evaluate \name with a mobile phone and a \wifi packet sniffer on a representative set of clandestine sensors and show a detection rate of $95.2\%$ and device classification rate of $100\%$ when the injected multi-modal event is human motion.
\end{itemize}

\section{Background}\label{sec:background}
\newchange{We provide an overview of that state-of-the-art approaches to detecting the presence of wireless sensors in spaces. We then formalize the notion of detecting whether a sensor is monitoring a particular area.}
\subsection{Detecting Wireless Sensors in Spaces}
The general approach to detecting wireless sensors relies on the notion that a device's wireless communication \newchange{unintentionally} leaks information in some \newchange{out-of-band channel}. \newchange{Recent works exploited these leaks to detect the presence of wireless, transmitting bugs\footnote{A \textit{bug} in this context refers to a hidden device spying on the user.} in a space~\cite{sathyamoorthy2014wireless, spyvsspy}.} The received power threshold and frequency range can be set according to a target set of wireless devices. For instance, to detect sensors that communicate over \wifi, a device would scan frequency ranges around $2.4$ GHz or $5$ GHz. In tuning the received power threshold, there is a direct trade-off between detection accuracy and false positives\cite{sathyamoorthy2014wireless}. If the threshold is too low, one may falsely attribute wireless signals from other devices in the space, like mobile phones, to bugs. On the other hand, a high threshold risks ignoring wireless bugs that are not within close proximity of the detector. As these detectors provide no semantic information about the detected signals, it is difficult to assume whether or not the observed signal is truly originating from a hidden bug\cite{spyvsspy}.

As wireless sensors transmit their information via packets, another technique to detect them uses packet sniffing.  Approaches like DewiCam \cite{cheng2018dewicam} sniff wireless packets and use their characteristics to train a classifier to identify whether or not a particular device is a camera. However, even if the type of device is determined, it may or may not be monitoring the user. If there is a camera monitoring the door of a house, it does not pose the same threat to a user's privacy as a camera that is monitoring the bedroom. Hence, even if we are able to detect what type of device is present in the space, it is difficult to characterize if its intention is adversarial. A direct way to identify whether a device poses a potential privacy threat is to determine whether or not it is actively monitoring the user.

\subsection{Detecting Sensors Monitoring a Space}\label{sec:causality} 
If a wireless sensor is monitoring someone in a physical space, the data that it captures is a function of the person's interaction with the space. For example, if someone moves into a space monitored by a motion detector, the sensor's control mechanism may be triggered and begin uploading relevant information to the cloud to be processed and forwarded (e.g., an alert to the device owner or downstream actuation). 
Similarly, the information recorded by a video camera captures variations \newchange{due to} motion within the captured scene--a characteristic exploited by prior research on detecting hidden cameras~\cite{nassi2019drones, liu2018detecting,wu2019you}. \newchange{To generalize across sensor modalities, we formalize the notion that if an auxiliary sensor observes and measures a user's interaction with their surroundings, we can identify whether the user's actions indicate a causal relationship with the hidden sensor's wireless traffic}. If such a relationship is found, then the sensor must be monitoring the user. 

\noindent\textbf{Detecting causality across sensor modalities.} 
\newchange{Given a target hidden sensor and access to its sensor data, we aim to establish causality between its time-series data and another sensor capturing the private space.} A popular method to study causal relationships between two series is Granger Causality \cite{granger1969investigating}. According to Granger Causality, if a series $X$ Granger-causes series $Y$, then past values of $X$ should contain information that helps predict $Y$ above and beyond the information contained in past values of $Y$ alone. Formally, if we have a series $Y$ as:
\begin{equation}
\label{eq:no-cause}
    y_t = a_0 + a_1*y_{t-1} + a_2*y_{t-2} + .... + a_n*y_{t-n},
\end{equation}
\noindent  and we augment this series with the series $X$ as follows:
\begin{equation}
\label{eq:cause}
    y_t = a_0 + a_1*y_{t-1} + .... + a_n*y_{t-n} + b_1*x_{t-1} + .... + b_m*x_{t-m},
\end{equation}
\noindent\textbf then $X$ Granger-causes $Y$ if and only if Equation \ref{eq:cause} gives a better prediction of $y_t$ than Equation \ref{eq:no-cause}. Here, $y_{t-k}$ are called lags of y and $x_{t-k}$ are called lags of x where $k \in [1, n]$. \older{However, several design challenges and goals lead up to establishing Granger Causality between a trusted sensor and a remote sensor in a generalized fashion.} 

\newer{In the following section, we discuss the system model and the design of \namenospace.}
\begin{figure*}[t]
\centering
\includegraphics[width=0.8\linewidth]{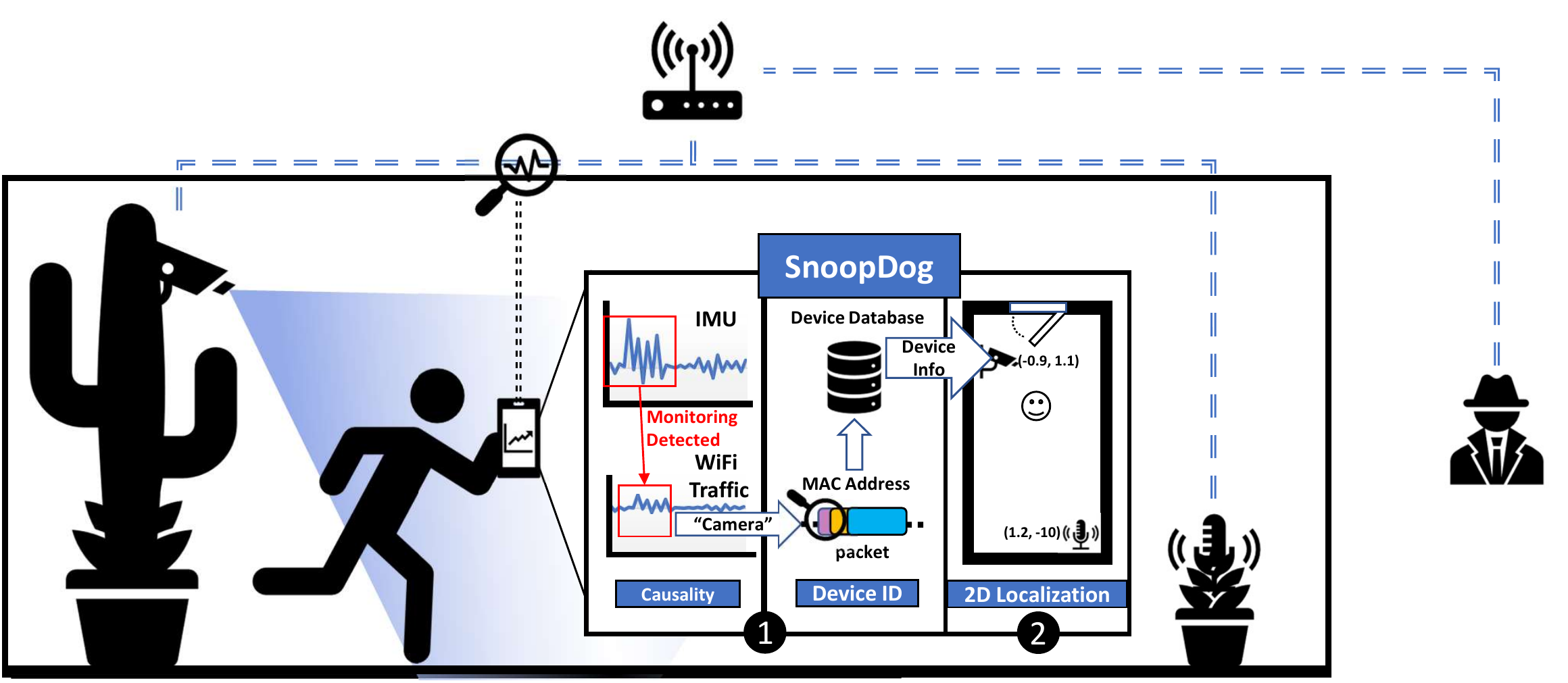}
\vspace{-0.08in}
\caption{Overview of SnoopDog framework. (1) The \name framework first identifies if a user is being monitored based on the cause-effect relationship between the values of a trusted sensor, e.g., an IMU,  and \wifi traffic patterns. It then inspects the associated packets and identifies the possible devices based on the physical (MAC) address. (2) Finally, \name localizes each device by leveraging directionality and sensor coverage.}
\vspace{-0.2in}
\label{fig:overview}
\end{figure*}
\section{\name Overview}\label{sec:design}
\newchange{We present the \namenospace's threat model assumptions prior to enumerating the system design.}
\subsection{System Model}
We consider a system model for \name where a user has access to a laptop or smartphone device with a network card that can enter monitor mode to sniff wireless packets over the same channel as one or more clandestine sensors. The system should further be equipped with a trusted set of \textit{ground truth} sensors to establish causality between the sensor values and the associated \wifi patterns from the clandestine wireless sensor(s)\footnote{We assume there may be additional, non-clandestine sensors that are monitoring the user. Such superfluous information is still informative, as the goal of this work is to detect all wireless sensors monitoring a user.}. These capabilities entail a set of assumptions. 

\noindent\textbf{\wifi sniffing assumptions.} We assume that the \wifi sniffer on the user's device can monitor the encrypted traffic streaming from the clandestine device. \name does not require any form of granted access to a particular network, i.e., \name should be able to sniff the device regardless of whether or not the network is closed or hidden. Unlike previous solutions, this implies that the user does not need to know the SSID or password of the network.

\noindent\textbf{Causality assumptions.} We assume that the user has a sufficient set of trusted ground truth sensors whose modalities are sensing any of the user's activities that would exhibit a causality with the \wifi encoding patterns of any clandestine wireless sensors. The notion of sufficient causality was formalized in Section~\ref{sec:background}. 

\subsection{Adversary Model}
\newchange{We focus on adversaries} whose goal is to remotely spy on a third-party occupant of a private space in real-time. \newchange{This model is consistent with other state-of-the-art methods for detecting hidden cameras~\cite{cheng2018dewicam,wampler2015information,nassi2019drones,liu2018detecting}, and is supported anecdotally by several cases where owners were live-streaming guests in private spaces, e.g., \cite{airbnb-news, jeong_griffiths_2019}. Further, many commercially available devices do not offer a local storage option for reasons of size, weight, power, and cost -- such is the case with six out of the popular thirteen devices we examined. Moreover, live-streaming offers a more practical and scalable solution from a management perspective}. Thus, we assume the adversary uses an arbitrary set of wireless, commercial-off-the-shelf (COTS) sensors that are tailored for clandestine placement. The communication between the attacker and sensor may be encrypted and placed on an arbitrary wireless frequency band. We further assume the adversary has deployed these clandestine sensors in a manner that is not apparently visible to the user within the space. We focus on an attacker utilizing devices that communicate over \wifi, as this is the most prevalent method of wireless communication for remote monitoring using commercial and consumer equipment\footnote{Although \name focuses on \wifi-connected devices, we discuss in Section~\ref{sec:limitations} how such a system could be generalized to other wireless communication standards and protocols.}. \newchange{An adversary may use one of the several techniques mentioned in Section \ref{sec:fool-snoop} to fool \namenospace, for example with cover traffic or local storage. Implementing these techniques can require modifying the device firmware or physically interfacing with a proxy device (e.g., RPi), thereby increasing the barrier-to-entry for potential attackers. Moreover, techniques such as cover traffic can add significant and undesirable network overhead, particularly for a large number of sensors.}
\vspace{-0.18in}
\subsection{Design Overview}
As depicted in Figure \ref{fig:overview}, \name detects and localizes a wireless sensor given access to a trusted sensor that can measure and quantify the ground truth in the modality that we are trying to detect. \name works in two phases. \circled{1} \textbf{Detecting and identifying snooping wireless sensors.} When a user first enters a new space, \name operates in a background mode to determine whether a user is being monitored based on the cause-effect relationship between the values of a trusted sensor (e.g., an on-body IMU) and \wifi traffic patterns. If the user wants to scan a room immediately, the background phase may be optionally skipped; alternatively, the background phase offers a low-overhead solution to bug detection. If a clandestine sensor is discovered, \name asks the user to perform a unique perturbation in the space to further ascertain the presence of a snooping sensor. The associated packets are then inspected to identify the possible device type based on the physical (MAC) address. \circled{2} \textbf{Snooping sensor localization.} In the second phase, \name utilizes a trial-based localization technique to identify the specific placement of the monitoring device. With the appropriate selection of ground truth sensor, that is, a device which can semantically capture at least a subset of the events captured by the snooping device, \name can detect clandestine wireless sensors of arbitrary modality. 

\section{Detecting and Identifying Snooping Wireless Sensors }\label{sec:cause-effect}
This section outlines the ability of \name to detect whether a clandestine sensor is actively snooping on a user. We describe the search space for wireless sensors, how to establish causality, how to generalize across modalities, and how to understand various sensors' wireless transmission. 
\vspace{-0.18in}
\subsection{Searching for Wireless Sensors}
The adversary can create a \wifi network and connect the snooping device to it. As a result, the hidden device can be present in any of the possible \wifi channels. Even though \name does not need access to these networks, it still needs to scan all \wifi frequencies and look for any devices transmitting on them. $2.4$ GHz and $5$ GHz are the most popular bands for \wifi networks, and as such, we focus on those particular bands, even though the \name scan region can be easily extended to include other ranges. 
During discovery, the \wifi Network Interface Card (NIC) scans through all channels sequentially to find available access points (APs) \cite{wu2009footprint, hu2015there}. Similarly, \name also scans through all the \wifi channels in monitor mode, but instead of looking for available APs, it looks for transmissions in those channels and creates a list of devices using the MAC address present in packet headers. As a result, \name does not need to be connected to any specific AP to operate. Even if a network is hidden, its transmissions can still be observed by monitoring the \wifi channel. Thus \name can detect devices on any \wifi network. Because devices may transmit data intermittently, \name continuously scans all \wifi channels and actively maintains an aggregate set of traffic data. Once the list of devices has been populated, \name then seeks to detect causality between user activity and data being transmitted from each device.

\subsection{Detecting Causality with User Activity}
Detecting the cause-effect relationship between the action of a user in a space and the data captured by a clandestine, wireless sensor requires access to two essential components: 1) a ground truth sensor to capture information about the user in the space and 2) a representation of the data collected by the clandestine sensor. While data packets transmitted by wireless sensors may be encrypted, the header information is not. This header information provides us with the MAC address and payload size of each transmitted packet. This data can be grouped and aggregated for all the packets within a time window and  provide information as to how much data was transmitted by each device within that period. Given a ground truth sensor, one can then identify causality between the ground truth sensor values and the patterns in the volume of data transmitted by each device in the space. In contrast to machine learning techniques, a causality approach allows \name to find the cause-effect relationship of arbitrary modality across any device that is transmitting causal data. \newchange{Because we are interested in the causality between two sensors, \name will utilize Granger Causality (described in Section~\ref{sec:background}).}


\subsection{Characterizing a Representative Set of Snooping Sensors}
In order to choose a set of ground truth sensors that can capture causality across any modality, we focus on generalizing across a representative set, including cameras, RF, and arbitrary sensors that report inferred (as opposed to raw) events. 


\noindent\textbf{Visual sensors.} Wireless cameras are typically encoded with a codec that recognizes underlying patterns in the frames of the video and utilizes this information for compression. 
One such codec is H.264 \cite{wenger2003h}. An encoder first encodes the video using the standard, and a decoder then reconstructs the original video with minor information loss. 

Standard temporal compression algorithms compress the video with 3 key frame-types, denoted I, P, and B frames. I frames (Intra-coded picture) hold complete image information, whereas P and B frames contain fractional image information, i.e., scene differences. As I frames are a complete image, they do not require any other frames to be decoded. P frames (Predicted picture) only contain changes in the image from previous frames. The information in a P frame is combined with the information of the I frame preceding it to obtain the resulting image. B (Bi-directionally predicted pictures) frames can construct the image from either direction using either changes from the I or P frames before them, changes from I and P frames after them, or interpolation between the I/P frames before and after them. B frames are most compressible, followed by P frames, and finally, I frames.

Hence, with increasing motion in the scene recorded by an IP camera, there will be an increase in the data that must be transmitted due to the increase in the number of P and B frames sent. Camera traffic will increase as the number of pixels being perturbed in the scene increases; similarly, traffic will decrease if the scene transitions to a stationary one. As such, if a human subject were to perform some motion in the scene, stop for enough time to let the camera traffic settle down, and then move again, it will result in a unique camera traffic pattern that corresponds to the user's motion. This cause-effect relationship between human motion and camera traffic can then be used to discover if a wireless IP camera is present in an occupied space. If there is no relationship between the camera traffic and user motion, then the camera is not monitoring the user.


\noindent\textbf{RF sensors.} Low cost, off-the-shelf millimeter-wave (mmWave) RF sensors are available that record the scene in the form of point-clouds. Recent works \cite{singh2019radhar, zhang2018real} have shown that these point clouds can be used to infer human activity. However, unlike a camera, a radar device is a point scatterer. Thus, at any given time, only certain points in the scene reflect back. Hence, with motion in the scene, the number of points captured in every frame by the sensor (radar) vary considerably. In an empty scene, the number of points captured by these sensors is fairly constant but varies as subjects move about the space. If such a sensor live-streams point-cloud data over \wifi, the payload size will vary over time with changes in the number of points captured in the scene by the sensor. Hence, the network traffic will fluctuate with the number of points that are being captured in the frame. As such, there exists a cause-effect relationship between the subject's motion and the device's traffic.  

\noindent\textbf{Acoustic sensors.} Another common type of bug used to snoop on people is a microphone. With the growth in personal home assistant devices such as the Google Home or Amazon Echo (Alexa)~\cite{kepuska2018next}, it is trivial for someone to buy and install such listening devices in their homes. Although they are typically triggered by a keyphrase such as ``Okay Google" or ``Alexa", there are ``Drop In" features that facilitate remote snooping. An adversary can also change the wake word of these devices to enable recording conversations of interest. Due to their compact form factor, they can be easily hidden. In such cases, these devices will also work like event-based clandestine sensors. Hence, services like \name that monitor traffic for change in network patterns and either correlate them with another sensor recording of the same modality or find a cause-effect relationship with the ground-truth can detect their presence using network sniffing \cite{kennedy2019can, wright2008spot}. Here, instead of the IMU, we use the microphone on the user's smartphone as the trusted ground-truth sensor. In section \ref{sec:discussion}-Q4, we discuss why it is challenging to detect and localize acoustic sensors that are continuously streaming.

\noindent\textbf{Wireless sensors that encode inferred events.}
Motion sensors do not transmit a continuous stream of information. Most off-the-shelf motion sensors are passive infrared (PIR) based. They measure the infrared (IR) light from objects in their field of view. Any change in this incoming IR light is inferred as motion. Instead of continuously transmitting, they send data to their cloud service for processing once triggered by motion. Thus, if a user moves around the room, stops, and moves again, there will be a unique cause-effect relationship between user motion and device traffic. Additionally, a camera can be programmed to continuously record video but only upload when a certain event occurs in the scene. These cameras behave like motion sensors and hence can be treated similarly. Virtual assistants also wait for trigger words to transmit a request to the associated cloud service, e.g., a user uttering the device name to activate it~\cite{kepuska2018next}.

\vspace{-0.05in}
\subsection{Device Identification via MAC Address}\label{sec:mac}

A MAC address is a universally unique ID assigned to the Network Interface Controller (NIC) for every networked device. It consists of 48 bits which are typically represented as 12 hexadecimal characters, i.e., \texttt{xx:xx:xx:xx:xx:xx}. The first 24 bits are the OUI (Organizationally Unique Identifier), which can uniquely identify a manufacturer or a vendor. 

The MAC address of the sender and the receiver are contained within each exchanged \wifi packet. More importantly, this information is not encrypted. As a result, \name can obtain the MAC address to look up the device vendor. While we acknowledge that the MAC address can be spoofed, this technique can still prove useful in the many cases where the adversary is a non-expert and thus has not spoofed the MAC address. Traffic fingerprinting techniques \cite{gao2010passive, crotti2007traffic, apthorpe2018keeping, zuo2019automatic, ortiz2019devicemien, meidan2017profiliot, miettinen2017iot} can also be used to overcome the shortcomings of MAC-based identification. Additionally, in case of MAC randomization or MAC spoofing, techniques such as the ones described in \cite{vanhoef2016mac} can be used to first track the traffic from a particular device and then perform cause-effect analysis on it.

\name contains a database with names and MAC addresses of known vendors that manufacture surveillance devices. As \name detects more sensors, we add them to the database.

\section{Snooping Sensor Localization}\label{sec:id-and-loc}




\SetKwFor{Loop}{Loop}{}{EndLoop}
\begin{algorithm}[t]
\DontPrintSemicolon 
\KwIn{
The sensor's $MAC$ address \\
\quad \quad \quad The $region$ of interest}
\KwOut{The sensor's location within the region}
$BBox \gets \emptyset$\;

$traversing \gets \textbf{BeginTraversingRegion(}region\textbf{)}$\;
\While{$traversing$}{
    $userloc \gets \textbf{DeadReckoningLocation()}$\;
    $inView \gets \textbf{GrangerCausality($MAC$)}$\;
    \If{inView}{
        $BBox \gets BBox \cup \{userloc\}$\;
    }
    $traversing \gets \textbf{SparseBBox(}BBox\textbf{)}$\;
}

\Loop{}{
    $MLE \gets \textbf{MostLikelySensorLocation(}region,  BBox\textbf{)}$\;
    \If{\textbf{SufficientBBox(}region, BBox\textbf{)}}{
        \Return{$(BBox,MLE)$}\;
    }
    $trialRegion = \textbf{GenerateTrial(}MLE, BBox\textbf{)}$\;
    $inView = \textbf{PerformTrial(}trialRegion\textbf{)}$\;
    \If{inView}{
        $BBox \gets trialRegion$
    }
    \Else{
        $BBox \gets BBox \setminus trialRegion$
    }
}

\caption{{\sc Localize} identifies the location of a particular snooping sensor in a defined region-of-interest}

\label{algo:localize}
\end{algorithm}

Algorithm~\ref{algo:localize} details the \textbf{trial-based localization} used by \name to infer sensor location.
In the case of multiple active sensors, this process can be repeated for each device.

\noindent\textbf{Setup.} Localization requires two input parameters: a region-of-interest to search over, and the snooping sensor's MAC address. 
To define the region-of-interest, we leverage \newer{Dead Reckoning \cite{patel,levi1996dead,beauregard2006pedestrian} for} indoor user localization. \newer{A dead reckoning mobile application\cite{patel} on a user's phone instructs} the user to walk the perimeter and capture the region boundary. 
Aside from identifying Granger causality in traffic patterns, the MAC address is also used to ensure an appropriate trial method for localization (e.g., via techniques discussed in Section~\ref{sec:mac} and~\cite{huang2019iot}). 


\subsection{Identifying Sensor Coverage}

Although the malicious sensor is known to monitor somewhere within the region-of-interest, it is unlikely to cover the entire region. Lines (1)-(8) narrow down the full  search space into a bounding box \emph{BBox} of the sensor's field-of-view.
To begin, a user is instructed to traverse the region (line 2). At regular time intervals, the user's location is captured, and the snooping sensor's traffic is monitored for causality. 
Using the Granger Causality technique described in Section \ref{sec:cause-effect}, a particular location is identified as either within or outside sensor coverage. This process continues until the bounding box is determined to have sufficient density for performing trial-based localization, depending on the coverage area size.


The remainder of Algorithm \ref{algo:localize} (lines 9-18) reduces the \emph{BBox} scope of sensor coverage via directional elimination. Repeated trials are performed to specifically target high-probability origins in order to either identify or eliminate likely sensor locations. Each round begins by solving for the most likely origin \emph{MLE} for the sensor (line 10). While this process could be performed randomly, utilizing physical information about the current bounding box can significantly reduce the number of necessary trial rounds. For example, if the bounding box shape can be reasonably fitted to a triangle, then the sensor is likely horizontal-facing and placed on a wall. On the other hand, an ellipsoid coverage area likely indicates a sensor placed on the ceiling or floor.

An iterative process then proceeds to reduce the area of possible sensor locations to a pre-defined threshold (e.g., 10\% of the region), upon which the bounding box and MLE are returned (line 11).
In each iteration, a \emph{directional} trial is conducted. \textbf{GenerateTrial}  identifies a suitable position and heading for the trial by selecting a point near the center of the bounding box and facing the MPE (line 12).
In our evaluation, we found distances of approximately 3 meters to be the maximum applicable distance for a trial.
The trial takes one of many forms; for an inertial sensor, a user faces the designated direction and waves an object (e.g., hand or shoe) 
closely in front of their chest while shielding this activity with their body from any sensor present behind them. To trigger a camera sensor, a laptop plays a video clip that randomly flashes the screen with different colors. For audio, a trigger sound is played, and so on. If the trial results increased the device traffic, the bounding box is reduced to areas within visible range (line 16); otherwise, those areas are removed (line 18), and the next iteration begins.

\vspace{-0.18in}
\subsection{Ensuring Sufficiently Reduced Region}

In order to provide a guarantee that this localization method will always result in a minimal bounding box that is sufficiently small (e.g., 10\% of the search region), a key assumption must be made: for any arbitrary bounding box, a trial can be identified which will eliminate a proper subset of the bounding box. In the case of Algorithm ~\ref{algo:localize}, this assumption can be reformed such that one can always construct a trial that eliminates at least a \emph{single} point contained within the bounding box set. Due to the directional nature of each trial, this can be achieved simply by conducting a trial that is positioned directly between two points within the bounding box, and facing directly towards one of the two points such that the other is obstructed. In the case of two points with large intermediate distances, a two-phase trial must be performed facing towards (and away from) each point, respectively.

Given the assumption that every trial can eliminate at least a single point from the bounding box set, guaranteeing that Algorithm \ref{algo:localize} will always reduce the region to a certain size is trivial. In the worst case, for a bounding box of \emph{n} points, \emph{n-1} trials must be performed. In practice, each trial can eliminate many points contained within the bounding box. Furthermore, by leveraging the most likely sensor location, one can reduce the search space significantly and with relatively few trials.

\section{Implementation}\label{sec:implementation}

This section presents the implementation details of \namenospace. We use readily available tools that are likely to be in a user's possession.

\subsection{Experimental Setup}
\noindent\textbf{\wifi Packet Sniffing:} The laptop's (Lenovo Thinkpad) network card enters monitor mode and uses Wireshark to capture all transmitted packets in the \wifi frequency band to aggregate traffic statistics for analysis. As it is not necessary to connect to a specific \wifi network to monitor traffic, \name can capture and identify clandestine wireless sensors across all \wifi traffic, even if they reside on a closed or hidden network. A smartphone can also be used instead of a laptop, but requires a rooted\cite{sun2015android} phone.

\noindent\textbf{Collecting User's Motion Data:} User's motion data is collected via the IMU present on the smartphone (Google Pixel 3). The smartphone is placed either in the user's hand or inside the user's pocket. 50 Hz accelerometer data is collected and used to study the cause-effect relationship between motion and sensor traffic. We collect data along each of the 3 axes and use them separately as if motion is present in only one direction, the other 2 axes contribute minimally to the analysis, and may instead serve as noise. The smartphone is also used to collect audio and localize the user in his/her surroundings.

\subsection{Detecting the Cause-Effect Relationship between User Motion and Hidden Devices}\label{sec:S5}

While sniffing the network, \name classifies the networked devices present into two categories: devices that transmit data continuously, and devices that have periodic or event-based transmission. 

\subsubsection{Wireless Sensors that Encode Raw Data}
Some representative sensors that continuously transmit variably encoded raw data include camera and RF sensors. 

\begin{figure}[!ht]
\centering
\includegraphics[width=0.7\linewidth]{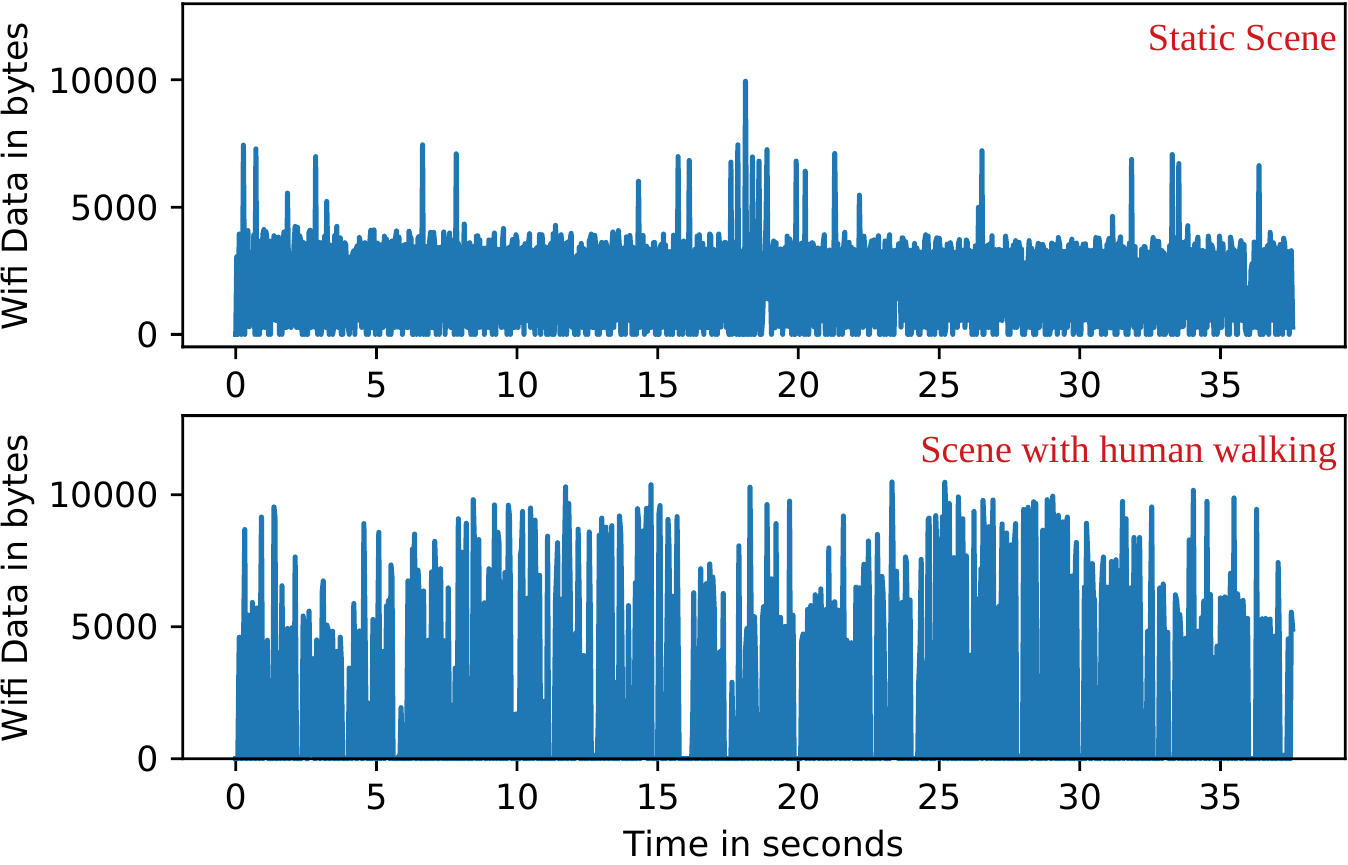}
\vspace{-0.08in}
\caption{\wifi traffic captured from a camera for a static scene and a scene where a human is walking around.}
\vspace{-0.12in}
\label{fig:cam_comparison}
\end{figure}

\begin{figure}[h]
\centering
\includegraphics[width=0.65\linewidth]{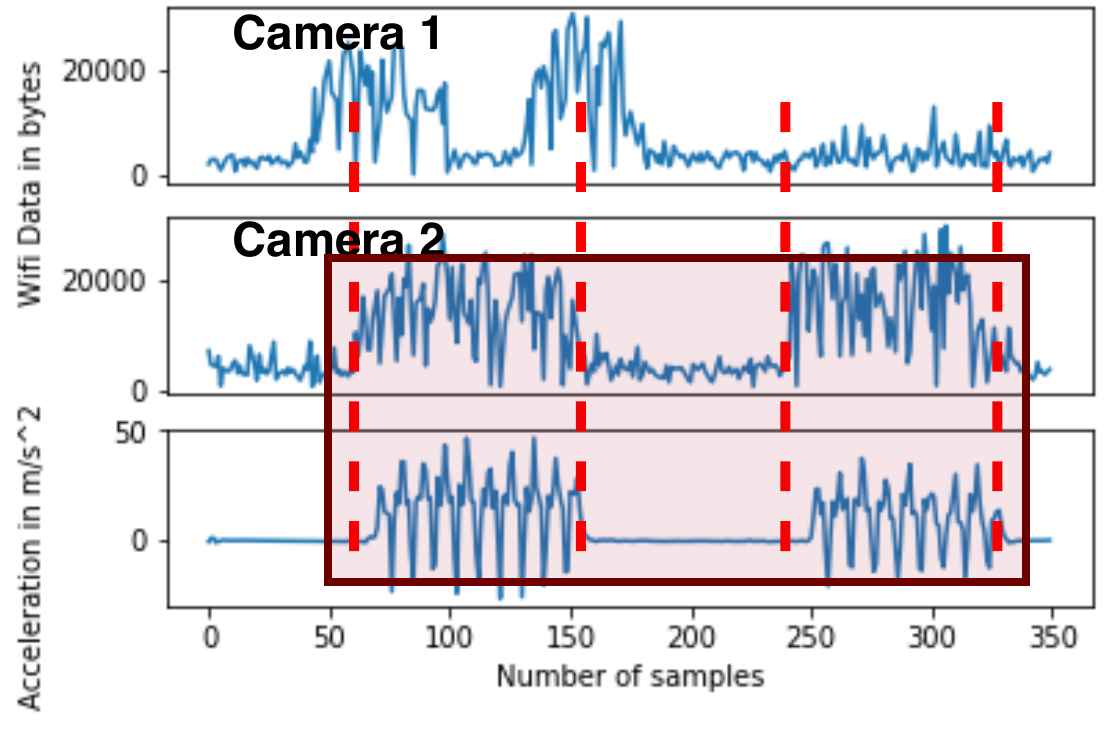}
\vspace{-0.08in}
\caption{\newchange{\wifi traffic of a snooping camera placed in the same space as the user (Cam. 2) and a non-snooping camera placed in a different space (Cam. 1) and its comparison with IMU data of the user being monitored in the scene.}}
\label{fig:cam_traffic}
\end{figure}

\noindent\textbf{Camera:} When a camera is monitoring a static scene, its traffic is fairly constant, as shown in Figure \ref{fig:cam_comparison}. As the scene is perturbed by human motion, the traffic changes rapidly. However, it is yet unclear whether human motion causes this variation. As soon as the user enters a new space, he or she can turn on \namenospace, which works in the background to correlate IMU data with \wifi traffic of the transmitting devices. As users walk in a space, the starting and stopping patterns of their motion are unique. This unique pattern creates a fingerprint on the camera traffic. Once \name is able to determine a cause-effect relationship between device traffic and user's motion, it alerts the user. To definitively ascertain the presence of a camera, \name asks the user to perform a stop-start-stop-start-stop (\textbf{S5}) motion as follows: 1) the user stays stationary for some time to allow the device traffic to stabilize. 2) The user performs jumping jacks at the current position. 3) The user stops again and waits for the device traffic to settle. 4) The user performs jumping jacks. 5) The user stops. The S5 motion causes a unique pattern to appear in the \wifi traffic as shown in Figure \ref{fig:cam_traffic} (Cam. 2). \newchange{}

The entire detection phase requires $35-45$ seconds. While the user is performing the above \textbf{S5} motion, \name sniffs the \wifi packets on the network and records the user's IMU acceleration. Figure \ref{fig:cam_traffic} plots the camera traffic after I-frame suppression and user accelerometer data while performing the \textbf{S5} motion. We observe that camera traffic is a function of human motion. When the human is static, the traffic is small, but when the human begins performing jumping jacks, the traffic rate increases. To prove that the accelerometer series indeed has an effect on the camera traffic, we leverage Granger Causality using the \texttt{statsmodel} package in Python. The null hypothesis of the Granger Causality Test is that the IMU series does not granger-causes the camera traffic series. Hence, if the p-value of our test is below the threshold of $0.08$, we can reject the null hypothesis and claim that the IMU series granger-causes the camera traffic series. We selected this p-value using the results obtained from the first camera. However, we evaluate our detection for all the other cameras and show that this p-value threshold is optimal for all the cameras.

\begin{figure}[h]
\centering
\includegraphics[width=0.85\linewidth]{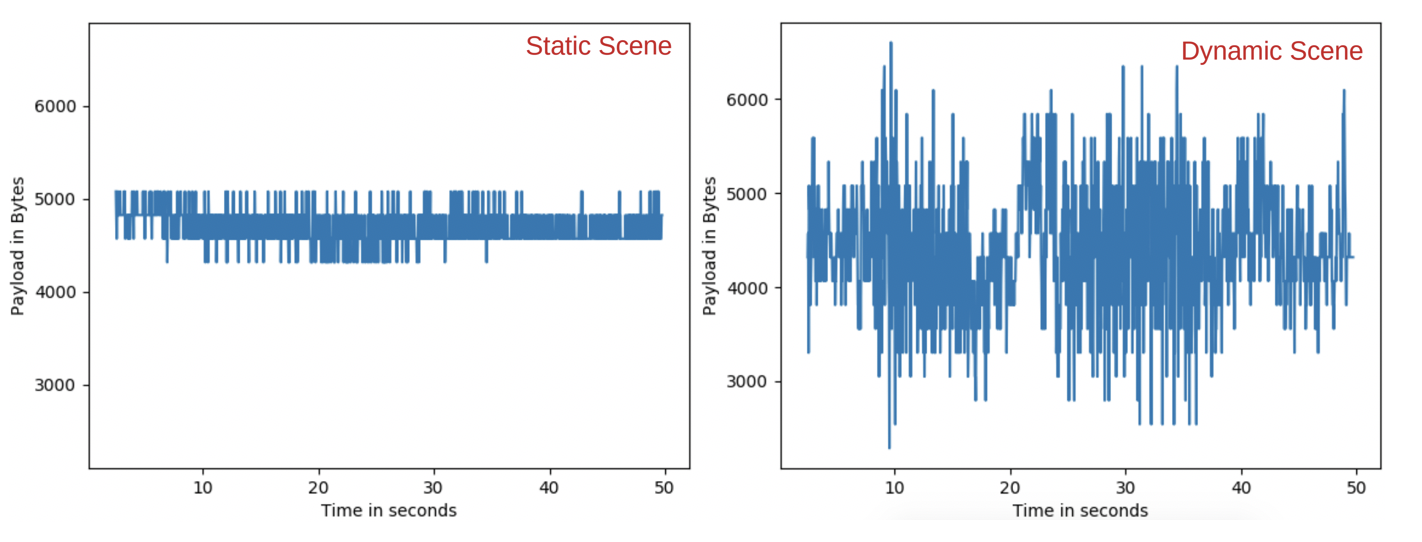}
\vspace{-0.08in}
\caption{Modeled \wifi traffic for an RF sensor in a static scene and one where a user performs our detection trial.}
\vspace{-0.25in}
\label{fig:rf_traffic}
\end{figure}

\noindent\textbf{RF sensor:} The detection process remains the same for RF as that of a camera. 
We use an off-the-shelf mmWave RF sensor from Texas Instruments, as shown in \cite{singh2019radhar}. We model the information obtained from the sensor as \wifi traffic. The modeled \wifi traffic from the RF sensor due to human motion is shown in Figure \ref{fig:rf_traffic}. Unlike a camera, RF sensors respond to a change in RF reflections from the scene.

As soon as motion occurs within the space, the traffic changes rapidly in response. This is because the points captured by the RF sensor vary with motion. If the traffic of some device which was static when there was no motion but changes rapidly when there is motion and goes back to being static when motion stops, it is an indicator that the device is monitoring user movement. To detect such devices, \name first monitors the traffic when the scene is static. It then asks the user to perform the \textbf{S5} motion in the space while \name monitors the traffic. As soon as the user is finished, the user should leave the space so that \name can monitor the traffic again and conclude the presence or absence of an RF sensor.

\subsubsection{Wireless Sensors that Encode Inferred Events}\label{sec:encode_timeout}
Sensors that encode inferred events transmit upon event detection. By examining network traffic, it is difficult to ascertain if the device is transmitting periodic data, like a temperature sensor, or transmitting inferred events like a motion sensor. 

\begin{figure}
\centering
\begin{subfigure}{.5\linewidth}
  \centering
  \includegraphics[width=.95\linewidth]{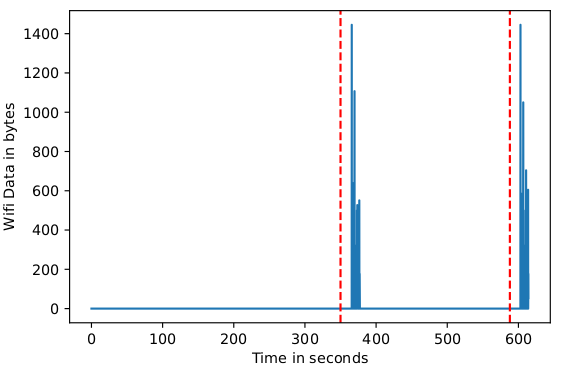}
  \caption{}
  \label{fig:sub1}
\end{subfigure}%
\begin{subfigure}{.5\linewidth}
  \centering
  \includegraphics[width=.95\linewidth]{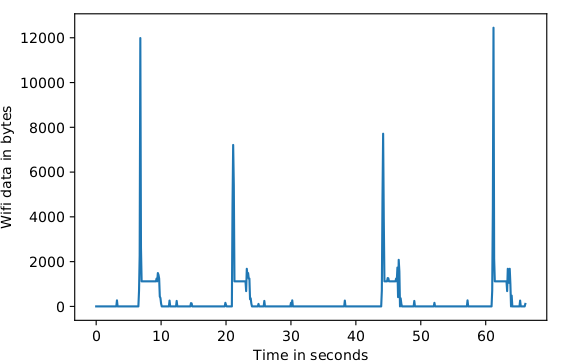}
  \caption{}
  \label{fig:sub2}
\end{subfigure}
\caption{(a) \wifi traffic of a motion sensor. The red-dotted line represents a motion event. (b) \wifi traffic of an Alexa device for the user repeating the same phrase 4 times.}
\vspace{-0.15in}
\label{fig:motion_traffic_alexa}
\end{figure}

\noindent\textbf{Motion Sensor: } Typical off-the-shelf motion sensors have a timeout to prevent continuous alerts. The motion sensor sends motion information to a cloud server, which in turn sends an alert to the snooping user's smartphone or performs an action like turning on lights. After sending an alert, the sensor waits for the timeout period before it looks for more events. This period is between 30 seconds and 3 minutes for most motion sensors. Similarly, there can be other sensors in the scene that have a timeout period between uploading events. To discover a device's timeout period, \name correlates user movements with device traffic. If two events are detected in the traffic of a device and the user was in motion during the time between the two events, this time is noted as the timeout period. \name uses its active phase to further improve the timeout estimation by asking the user to move around the space until two events are detected in the device's network traffic. \name asks the user to move around the space, leave the space for the timeout period, and then move around the space again. After that, the user moves out from the space and then waits for the timeout period to end. If \name detects traffic by the device around the same time the user moved and none when the user is not moving, it concludes that the traffic of the device is caused by user movement. This process can be repeated to increase the confidence of detection. In Figure \ref{fig:motion_traffic_alexa}, we move around the room and notice that the \wifi traffic from the motion sensor responds to these motion events. Since this traffic is discrete, we cannot perform time-series Granger causality analysis. Instead, we perform an activity and track network response. To detect the presence of a motion sensor, we ask the user to move around the room, wait for the timeout period, and move around again. \name scans all device traffic within a period of $5$ seconds after the motion to determine which device responds to user motion. If the device has traffic activity after the user moved, then the device is inferring events from the user motion.

\noindent\textbf{Audio snooping: } \name records user conversations in the background and monitors the network traffic. If the occurrence of a certain phrase or a word causes the traffic of a device to change, \name asks the user to repeat it until it can establish a causality between the occurrence of that phrase and the traffic of the device. Once \name knows the ``wake word" for the acoustic home-assistant device, it repeats the recording several times while monitoring the device traffic to increase the confidence level of detection.

In our implementation, we used an Amazon Echo and Echo Dot whose wake word was ``Alexa" \newchange{and ``Computer" and a Google Home Mini with the wake phrase ``Hey Google".} \older{In Figure \ref{fig:motion_traffic_alexa}, we say the phrase ``Alexa, what's the time right now?" four times and observe four distinct events in the device traffic. In 35 trials with different phrases, \name was able to detect causality $100\%$ of the time. Additionally, we show the variations in device traffic for Alexa Drop-In features when the Drop-In is active and when it is not active in Appendix \ref{sec:drop_in}.} 

\subsubsection{Device ID via MAC Address Lookup}

\name checks its database for a match of OUI in the device's MAC address. If present, \name can inform the user with higher confidence that the device is indeed a surveillance device. Otherwise, it is added to the database and identified as a clandestine sensor.

\subsection{Device Localization}

\name uses dead reckoning~\cite{ojeda2007personal} and asks the user to walk around the perimeter of the room to create rough map of the room. Next, the user performs a detection trial at various locations in the room. More trials lead to better localization. At every location, \name tries to establish a cause-effect relationship with the device traffic. Regions with no cause-effect relationship are eliminated.

\noindent\textbf{IP Camera:} The traffic generated by a camera monitoring a scene will increase when the scene is dynamic. To exploit this, we first monitor the traffic of the device identified as a camera for $30$ seconds over a static scene. Each trial consists of standing in a particular location (e.g., the middle of the scene), pointing a laptop in a particular direction, and playing a video that rapidly changes the colors on the screen of the laptop for $30$ seconds. This process is then repeated in different directions. If the camera is able to monitor the laptop screen, its data rate during that period will be higher. On the other hand, if the laptop screen is not visible, the camera's traffic rate will be similar to the static scene. We can eliminate a fraction of the space where no activity is detected and repeat the process for the remaining region. In this way, we narrow down the possible region where a camera is located. We give a step by step walk-through of this process in section \ref{sec:evaluation}.

\noindent\textbf{RF sensor:} RF sensor localization is similar to that of a camera. However, since RF sensors cannot detect the flickering screen of the laptop, we use human movement. \name asks the user to stand in the middle of the space and wave their arm up and down rapidly in front of them while shielding this motion from the other side of the space with their back. If the RF device traffic does not respond to these stimuli when performed on one side but responds to it on the other side, we can eliminate that space.

\noindent\textbf{Motion Sensor:} Motion sensors are triggered by motion in front of them. \name first identifies the motion detector timeout (refer section \ref{sec:encode_timeout}), and then asks the user to stand in the middle of the room before the timeout expires. After timeout expiry, they are asked to move their hand in front of them while shielding it from the other side with their body. 

\noindent\textbf{Acoustic (Audio) sensors:} \name records the wake word of the device and asks the user to move around the room while this sound is repeatedly played from the smartphone app. If the user walks around the room but does not find any place where there the traffic of the device changes, we increase the volume and repeat the experiment. On the other hand, if the sound played at every point in the room causes the traffic of the device to vary, we decrease the volume and repeat the experiment. Finally, we identify areas where the sound causes network response and areas where it does not. We continue to reduce the volume of the device until the search space has been sufficiently reduced\footnote{A walk-through of this process is provided in section \ref{sec:audio_loc} of the Appendix.}.

\vspace{-0.18in}
\section{Evaluation}\label{sec:evaluation}

We evaluated \name on a set of \newchange{sensors from well-known brands as well as best-selling sensors on Amazon. These are listed below in Table \ref{tab:list-sensors}.}

\begin{table}[h]
\centering
\resizebox{0.98\columnwidth}{!}{%
\begin{tabular}{|l|l|l|}
\multicolumn{1}{|c|}{\textbf{Name}} & \multicolumn{1}{c|}{\textbf{Type}} & \multicolumn{1}{c|}{\textbf{Cost}} \\ \hline
Kamtron                             & Camera                             & \$39.99                            \\
Panasonic (HomeHawk)                & Camera                             & \$77.64                            \\
Wansview                            & Camera                             & \$29.99                            \\
Arlo (NetGear)                      & Camera                             & \$107.50                           \\
Victure                             & Camera                             & \$35.99                            \\
Foscam                              & Camera                             & \$49.99                            \\
Ring (Amazon)                       & Camera                             & \$59.99                            \\
Amazon Echo Dot                     & Home Assistant                     & \$29.99                              \\
Amazon Echo                         & Home Assistant                     & \$99.99                            \\
Google Home Mini                    & Home Assistant                     & \$39.99                            \\
Kangaroo Home                       & Motion Sensor                      & \$12.95                            \\
Samsung Smart Things                & Motion Sensor                      & \$24.99                            \\
TI IWR1443                          & RF Sensor                          & \$299.99                          
\end{tabular}%
}
\caption{\newchange{List of snooping sensors evaluated upon}}
\label{tab:list-sensors}
\end{table}

\subsection{Sensors that Encode Raw Data}

\noindent\textbf{Wireless IP Cameras.} For Granger causality analysis, we lag the first series by one element at a time and observe what value of the lag results in the lowest p-value. Cameras have a delay between when the scene changes and when the data is visible to the adversary. We found that this delay can vary between a few milliseconds to up to 4 seconds. If the adversary is using a tape delay in transmission, we can perform this analysis over a longer delay period. 
Assuming symmetrical delay, \name sniffs the packets during the first half of the transmission; we choose a lag value of 2 seconds.


We evaluated our detection on 7 cameras. All of them use H.264/MPEG-4 codecs which are the most popular codecs used for IP cameras. We performed 131 trials on 2 different users\footnote{The data is collected from the authors and hence does not require IRB approval.} to evaluate the detection accuracy. The results of our experiments are presented in table \ref{tab:camera_eval}. To improve the detection accuracy and confidence of detection, a user can perform the detection trial several times and take a majority vote. The detection works well even when a portion of the human body is occluded by objects such as a table.

\begin{table}[h]
\centering
\resizebox{0.7\columnwidth}{!}{%
\begin{tabular}{|c|c|c|c|}
\textbf{Camera} & \textbf{Trials} & \textbf{Successful} & \textbf{Accuracy} \\
\hline
Panasonic       & 15              & 14                  & 93.33\%             \\
Arlo (Netgear)  & 10              & 10                  & 100\%             \\
Ring (Amazon)   & 10              & 9                   & 90\%             \\
Foscam          & 15              & 15                  & 100\%             \\
Wansview        & 30              & 29                  & 96.6\%             \\
Kamtron         & 25              & 21                  & 84\%              \\
Victure         & 26              & 26                  & 100\%             \\
\hline
\textbf{Total}  & \textbf{131}     & \textbf{124}         & \textbf{94.65\%}    
\end{tabular}
}
\caption{\newchange{Evaluation results for camera detection}}
\label{tab:camera_eval}
\end{table}

\noindent\textbf{RF sensors.}
We use a TI mmWave IWR1443 to evaluate the performance of \namenospace. In 20 experiments, \name was able to detect RF sensor's presence every time.

\subsection{Sensors Encoding Inferred Events}

\noindent\textbf{Motion Sensors.} We evaluated on an off-the-shelf motion sensor from Kangaroo Security and a smart-things motion sensor from Samsung. \newer{The smart-things sensors are a special case as these sensors use Z-Wave and ZigBee to communicate with a smart-things hub which in turn sends the information over \wifi. As a result, \name can sniff the traffic of this hub and establish causality. However, if there are multiple devices connected to the same hub, \name will not be able to detect them.} We performed 25 trials, and \name was able to detect the motion sensors every time except for 3 trials. We suspect that this was caused because the devices send some sort of ``status" messages to their respective cloud service which result in events in the sniffed traffic that throw the detection off.

\noindent\newer{\textbf{Smart-home Assistants (Audio Sensors).} In Figure \ref{fig:motion_traffic_alexa}, we say the phrase ``Alexa, what's the time right now?" four times and observe four distinct events in the device traffic. In 35 trials with different phrases, \name was able to detect causality $100\%$ of the time. Additionally, we show the variations in device traffic for Alexa Drop-In features when the Drop-In is active and when it is not active in Appendix \ref{sec:drop_in}.} 


\subsection{Quantifying False Positives and Negatives}\label{sec:false-positive}
\newchange{\noindent\textbf{Overall false positive rate.} We perform false positive analysis for two separate cases. In case one, we run cause-effect analysis for every device whose traffic has been captured (a network can have devices such as phones, laptops, and other sensors in addition to snooping devices) and find out if a device is falsely classified to be snooping on the user. In case two, we perform our cause-effect analysis on a camera that is not looking at the scene to see whether it will produce a false positive.}

\newchange{For case one, our experimental test-bed consists of devices monitoring the user such as a camera, motion sensor, home assistant and RF-sensor. We also include innocuous devices usually present in a scene such as smartphones, laptops, \wifi-routers, smart-lights, and embedded devices such as raspberry-pi. We ran this analysis for 50 trials. We found $2$ false positives for Active Phase and $9$ false positives for Background Phase.}

\newchange{We divide case 2 into two separate parts. In the first part, a camera is placed in another room (where the user who's running \name is not present). Another person is moving freely in the room. In the second part, the person in the other room is sitting on a desk and working while the camera is monitoring them. We show the results of our experiments in Table \ref{tab:fpa}. We perform these trial for both background detection and active detection. The one detection for background phase is because the walking patterns of the user and the person in the other room were similar. In such cases, a user can turn off the light in their room and see if the traffic of the device drops. If it does, it means the camera is in their room.}

\begin{table}[h]
\centering
\resizebox{0.98\columnwidth}{!}{%
\begin{tabular}{|l|l|l|l|}
\multicolumn{1}{|c|}{\multirow{2}{*}{Case}} & \multicolumn{1}{c|}{\multirow{2}{*}{Total Trials}} & \multicolumn{2}{c|}{False Positives}                               \\
\multicolumn{1}{|c|}{}                      & \multicolumn{1}{c|}{}                             & \multicolumn{1}{c|}{Background} & \multicolumn{1}{c|}{Active (S5)} \\ \hline
(1) User moving                             & 15                                                & 1 (6.66\%)                      & 0 (0.00\%)                       \\
(2) User working on a desk                  & 10                                                & 0 (0.00\%)                      & 0 (0.00\%)                      
\end{tabular}%
}
\caption{False-positives for a camera in a separate room}
\label{tab:fpa}
\end{table}

\begin{figure}[h]
\centering
\includegraphics[width=0.6\linewidth]{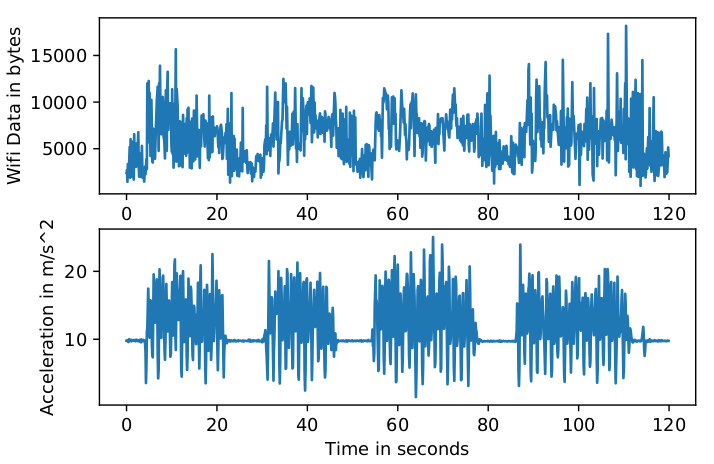}
\vspace{-0.08in}
\caption{Camera traffic and its comparison with IMU data}
\vspace{-0.2in}
\label{fig:background_fn}
\end{figure}

\noindent\textbf{False negatives for background detection.} \namenospace's background detection mechanism is most susceptible to false negatives since it relies on natural human motion rather than active command and response patterns. Figure \ref{fig:background_fn} shows camera traffic and IMU data for a user walking randomly in a room. As long as the user stops at some point within the detection period, \name is able to detect hidden devices. However, if the user is continuously stationary or continuously in motion, \name fails to detect the presence of clandestine sensors.

\subsection{Snooping Sensor Localization}\label{sec:localization_eval}

\begin{figure}[h]
\centering
\includegraphics[width=0.9\linewidth]{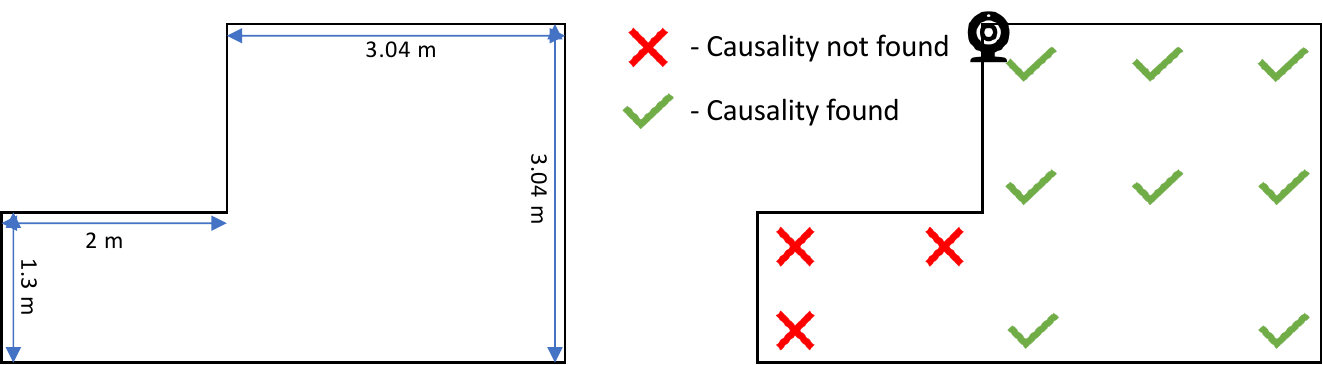}
\vspace{-0.08in}
\caption{Lab dimensions and results of the detection trials.}
\vspace{-0.1in}
\label{fig:lab-dimensions}
\end{figure}

We evaluated \name for 4 different spaces with different sensor placements. The accuracy of localization in all of these cases depends on the user's requirements. The user can perform more trials to reduce the probable region where the sensor is placed. We use an example to demonstrate how the \name localization algorithm works. To perform our localization, we chose a room as shown in Figure \ref{fig:lab-dimensions}. The camera is placed at a corner of the room. We begin by performing our \textbf{S5} detection trials in different parts of the room. The location and results of our trials are shown. Based on these observations, we know that the camera is present somewhere in the square region of the room and hence, we eliminate the other part and start our trial-based localization.

We stand in the middle of the probable space and hold a laptop such that the screen is pointing in one direction. Then we turn to the other side and repeat the same experiment. We observe that there is a significant ($>$150\%) increase in the camera data rate when the laptop is pointed towards the left side. When pointed to the right, the data rate remains similar to that of an empty room. Thus we eliminate the right portion of the room from the probable area. We again stand in the middle of the leftover space and repeat the experiments until we achieve a sufficiently reduced space.

\begin{figure}[t]
\centering
\includegraphics[width=\linewidth]{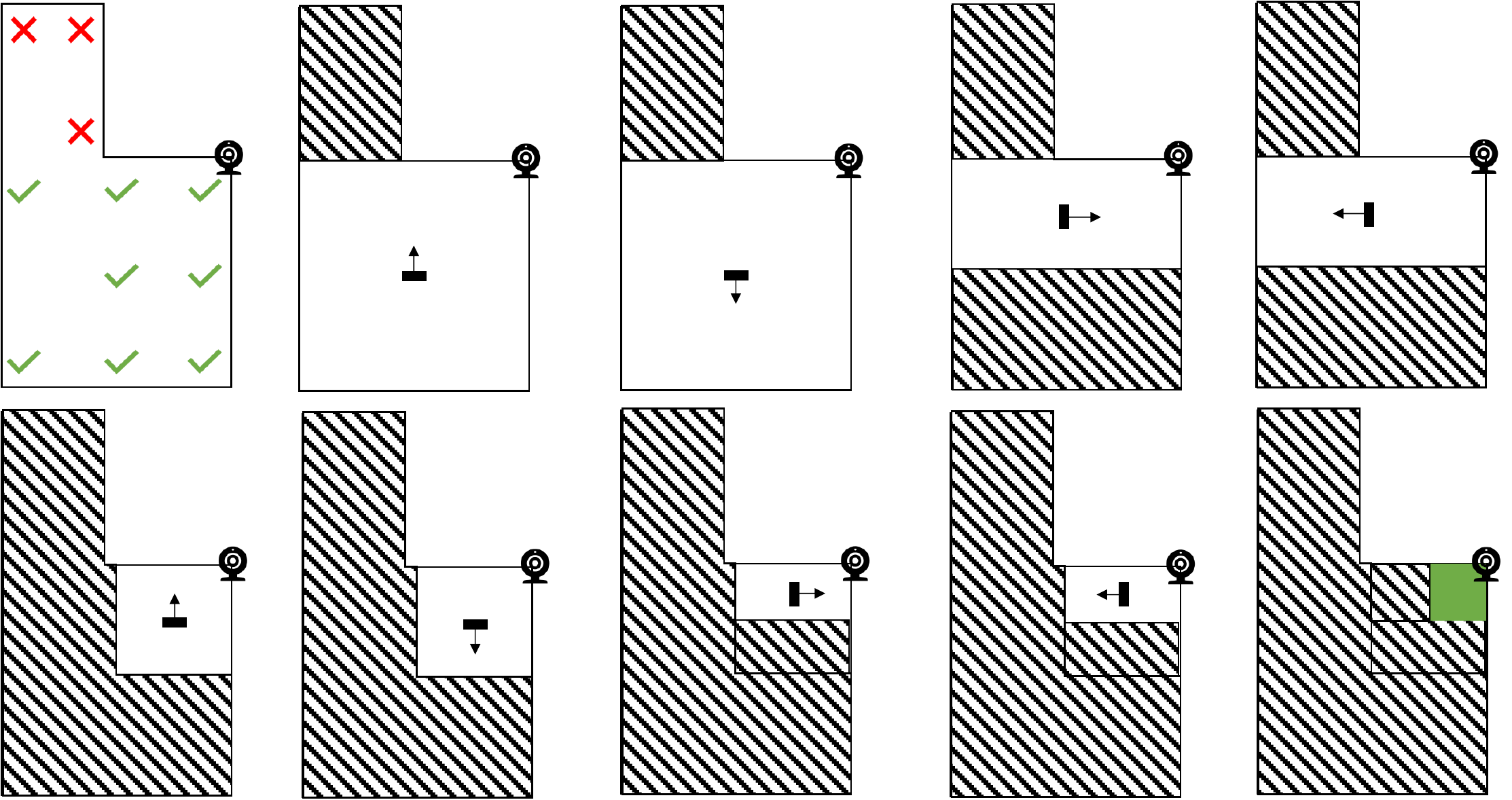}
\vspace{-0.15in}
\caption{A walk-through of the trial-based localization algorithm in the laboratory environment in Figure~\ref{fig:lab-dimensions}. The arrows represent the direction the laptop screen was facing.}
\vspace{-0.22in}
\label{fig:overview_loc}
\end{figure}

\noindent\textbf{Audio-based localization:} A similar elimination-based localization for audio sensors is described in Appendix \ref{sec:audio_loc}.

\subsection{Overhead Analysis}
\noindent\textbf{Time:} Sensor detection can happen in the background with minimal user intervention. However, this will take some time. In situations where a user wants to immediately know if he/she is being spied on by a sensor (such as when entering a changing room), they can directly begin the active phase where they will perform the \textbf{S5} motion. It takes about $40$ seconds to perform active detection. For localization, each trial can take $30$ seconds. Since the localization space reduction is determined by the user, he/she can perform the trial any number of times. If the total number of trials is $n$, the overhead will be about $30n$ seconds.

\noindent\textbf{User effort:} If the detection occurs in the background, there is no overhead in terms of user involvement. However, both active phase and localization require user effort. In case the user is suffering from physical disabilities, he/she may find it hard to follow through these steps. 

\noindent\newchange{\textbf{\wifi Channel Hopping:} In detection mode, \name must hop channels across all frequencies as an adversary may have configured devices on different \wifi frequency bands. A transmitting device might inadvertently evade detection if it is hopping while \name is scanning. Hence, we perform multiple scans of the entire \wifi spectrum to reduce misdetections. We would ideally monitor all \wifi channels simultaneously, but this is infeasible with a single WiFi radio. Multiple \wifi radios will reduce the search space per radio at additional cost and degraded portability.}



\subsection{Effects of Range on Detection}\label{sec:range-effect}

Detection depends on the range of a sensor. The camera range is the depth which it can record meaningful information from the scene. If the user stands too far from the camera, their motion may not produce a significant variation in its traffic. For motion sensors and RF sensors, the range is specified by the vendor. The motion sensor that we used has a range of $15$ feet ($4.6$ m) and the RF sensor we used had a range of $4$ m. For Amazon Echo, the range is a function of the loudness of sound. For cameras, we perform our detection trials at different distances to find out the range.

We perform the \textbf{S5} motion in front of a camera at $1$ m and gradually increase the distance. We find that as we move away from the camera, the changes in its traffic decrease. For the Wansview camera, the distance at which traffic changes are not enough to establish causality is the smallest at about $3$ m.

\vspace{-0.18in}
\newer{\subsection{Failure Analysis}}
\noindent During our evaluation, \namenospace's detection failed during some trials for both cameras and motion sensors. In this section, we will describe and discuss those cases.

\noindent\newer{\textbf{Camera.} \name failed to detect hidden cameras when the camera video froze (no transmission) for a small duration during our detection trial. Another cause of miss-detection were additional users in the scene since the network traffic due to their motion cannot be filtered out. Also, when the user was present far from the camera (refer Section \ref{sec:range-effect}), the traffic changes due to the motion of the user were not enough to establish Granger causality.}

\noindent\newer{\textbf{Motion Sensor.} Motion sensors sometimes send status messages to the cloud service that they are connected to. These transmissions happen in the form of a short burst. Since there was no motion before these transmissions, they result in \name failing to establish causality. While we cannot say with certainty what each of these transmissions are, we saw multiple such transmissions when the sensor was started, during update, and sometimes even in between two motion events. A user can also query the sensor status from their phone to which the sensor has to respond. \name failed to detect causality when these transmissions were present in the network traffic of the motion sensor.}
\section{Techniques to fool \name }\label{sec:fool-snoop}

In this section, we discuss how an adversary can fool \namenospace.
\subsection{No Encoding or Data Padding}
\begin{figure}[h]
\centering
\includegraphics[width=\linewidth]{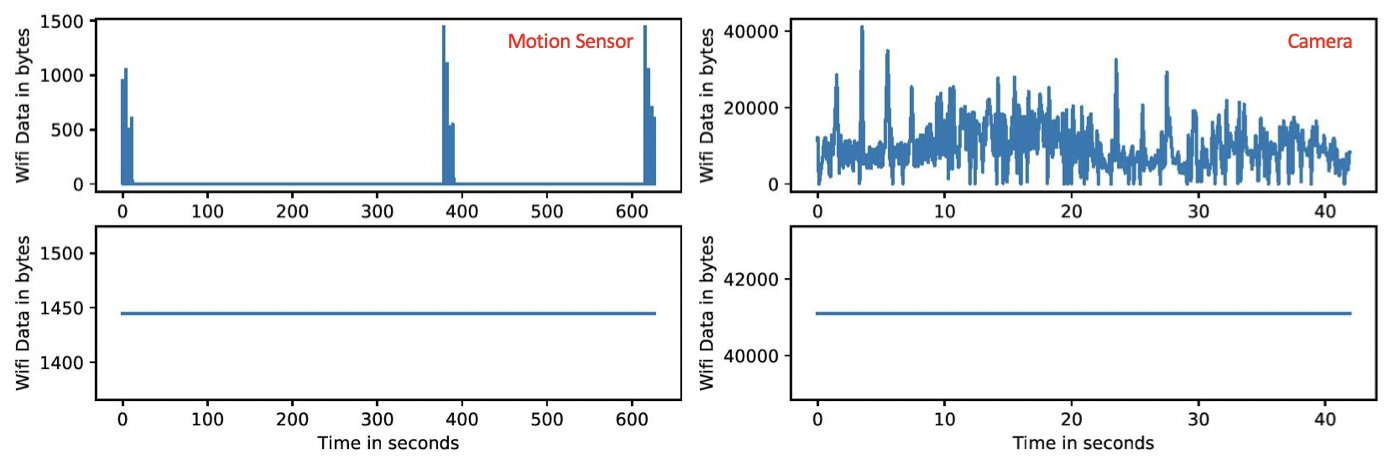}
\caption{Padding the motion sensor and the camera traffic}
\label{fig:padding_camera_motion}
\end{figure}

\name uses the relationship between encoding schemes and ground truth to find out if there is a device which is monitoring the user. Hence, to fool \namenospace, the sensors can either send un-encoded raw data or they can pad the encoded data to make the data rate constant. Cameras can either pad their traffic or they can send un-encoded images frames. Since sending images will put a large overhead on the network bandwidth, padding the traffic \cite{apthorpe2018keeping} is a better idea. We pad the camera traffic with random payload in Figure~\ref{fig:padding_camera_motion}. Since \name cannot see what's inside the payload, it can be anything. The device can even send labels in the payload that help the server decide if this is a valid packet or fake data generated to fool detection. Also in Figure~\ref{fig:padding_camera_motion}, we pad the traffic of a motion sensor to make it appear like a constantly transmitting device with no variation in traffic in response to user's motion.

For RF sensors, one can find out the maximum number of points it can output and then always pad the information so that we are transmitting the maximum number of points allowed. These extra points could all be zeros which would make it easier to filter them out on the server side.

\subsection{Adding Random Noise to the Data}

Another way to fool \name is by injecting noise into the device's wireless traffic at random intervals for some time window. Since \name utilizes the change in device traffic to ascertain a cause-effect relationship, the variations caused by injecting random noise are able to fool the detection.

\begin{figure}[h]
\centering
\includegraphics[width=0.8\linewidth]{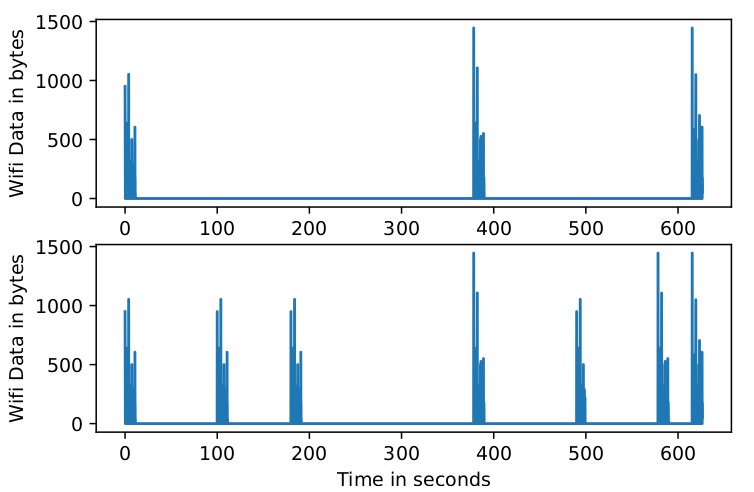}
\caption{Injecting noise in the traffic of a motion sensor to fool \name}
\label{fig:random_motion}
\end{figure}

Devices that do not transmit continuously can randomly send information that creates a pattern similar to their inferred event traffic. This way they can keep sending their information which is hidden within random traffic. We add random noise which appears like regular traffic for a motion sensor in Figure \ref{fig:random_motion}. 

\subsection{Constantly Vary the Resolution of the Data Being Transmitted}

For devices like camera, there are several video resolutions that an adversary can choose. The higher the resolution, the better the video quality is. However, if an adversary chooses a scheme where the video resolution is constantly varying, it will cause random changes in the network traffic. Hence, even if the user's motion is causing changes to the traffic, it is overpowered by the changes in network traffic due to a variation in resolution.

For RF sensors, they can vary the number of maximum points that they transmit continuously to achieve a similar effect.

\subsection{Adding a tape/broadcast delay to the transmissions}
An adversary can add a tape delay to the sensor transmissions, i.e. intentionally adding a delay between when something was recorded and when it was transmitted. Since, we are only looking for causality within a small time window, a high tape delay will be able to fool \name. However, given enough storage capacity and time, it is possible for \name to scan the entire recording to look for cause-effect relationship with user motion. But for large tape delays, this is not practical. 
\newer{\section{Limitations}\label{sec:limitations}}

\noindent\newer{\textbf{\textit{1: Only limited to VBR devices.}} Although \name can detect a wide variety of commonly available sensors, it cannot detect \textit{any} wireless sensor monitoring the user. For a sensor to be detectable by \namenospace, the traffic must be encoded with a Variable Bit Rate (VBR) algorithm and the data recorded by the sensor must change in response to user perturbation which can be recorded by a ground truth sensor. That said, most surveillance devices such as cameras, motion sensors and smart-home assistants today fall into this category, and thus we believe \name can serve as a valid defense.} 

\noindent\newer{\textbf{\textit{2: A technically capable adversary can fool \name if they know about its existence.}} If the adversary suspects \name is in use, they can use one of the techniques listed in Section~\ref{sec:fool-snoop}. They can also use channel hopping or MAC randomization. We have not evaluated \name for any of the above techniques.}

\noindent\newer{\textbf{\textit{3: Evaluation is limited to \wifi devices and devices who route their traffic through a \wifi-hub only.}} We have evaluated \name for \wifi-connected devices only. For future work, this framework can be evaluated for other popular wireless communication standards. \name can be extended to standards like Zigbee \cite{kinney2003zigbee}, Z-Wave\cite{yassein2016smart}, and Bluetooth\cite{muller2001bluetooth, haartsen2003bluetooth} as long as we have the following: 1) A receiver that can scan their probable frequencies and sniff their packets to find if any devices are transmitting and 2) the ability to find unique device IDs from packet headers and distinguishing header information from payload size. While capturing Zigbee/Z-Wave packets will require additional hardware, recent works have shown that it is possible for a \wifi radio to perform cross-technology communication. \cite{li2017webee, kim2015freebee}}
\section{Discussion}\label{sec:discussion}

\noindent\textbf{\textit{Q1: What is the usability of \namenospace?}} We envision \name to be implemented as an app on either a smartphone or a smartwatch (or a combination of the two). This means an end-user will not need any prior knowledge about causality and coverage of a device to use it. \name will continuously work in the background to look for a cause-effect relationship between a user's actions and device traffic. It will then guide a user step-by-step through the entire localization procedure. Since an adversary can place a sensor at any time (e.g.,when a user checks in a room, searches for devices, finds none and then leaves for dinner after which the adversary places the spying device), \name will still find it because it continuously works in the background. This will not cause any overhead in terms of user involvement.


\noindent\textbf{\textit{Q2: How can false positives be reduced?}} For false positive to occur during active detection, the device's traffic needs to map directly to the \textbf{S5} motion during the active phase and user's motion during the background phase, which is unlikely. If there happens to be another camera in an adjacent space monitoring another user who is performing the detection trial within the same time window as the first user, it will trigger a false detection. However, the probability of this happening is low. Nevertheless, it remains a possibility, and mitigating such instances are highly desirable.

Simple strategies can significantly reduce the chances of false positives. First, during the initial monitoring phase for wireless devices, any periodic trends in traffic patterns can be noted; the detector trial should ensure its periods are not synchronous with such periodicity. Furthermore, the detection process can be done multiple times with varying and erratic period lengths. This will drastically decrease the chances of a false positive, as a device would have to coincidentally follow this effectively random traffic pattern. Finally, the entire process itself can be performed repeatedly; each iteration compounds the decrease in false positive rate, such that it eventually reduces to a statistical impossibility.



\noindent\textbf{\textit{Q3: Are there alternative approaches to causality?}} One alternative approach to detecting snooping sensors is correlation. However, correlation does not imply causation. If we have a sensor that measures the ground truth in the modality we want to detect, we need to use causality analysis. For example, it takes the camera some time to process the information and send it over to the server. So if we capture human motion with an IMU, the camera traffic will lag the IMU time series. This is correctly captured by causality analysis but not by correlation. However, if instead of using a sensor to measure the ground truth, we use another sensor that can capture the same modality that we are trying to detect, we can use correlation because if both the devices are capturing the same event, their traffic should show similar trends. Future work can also explore the efficacy of data-driven approaches such as deep learning for time series classification.



\noindent\textbf{\textit{Q4: Can we detect continuously streaming audio bugs?}}\label{sec:why_not_audio} There are two ways to encode audio, either constant bit rate (CBR) or variable bit rate (VBR). VBR techniques make use of similarity in sound, such as prolonged silence, to reduce the amount of data required for encoding. In contrast, CBR always encodes with the same number of bits. Many off-the-shelf audio recorders and audio streaming apps use CBR. Since \name only has access to the payload size of a packet, there must be variation in the payload to determine causality. Hence, \name cannot detect CBR audio bugs.

\noindent\newchange{\textbf{\textit{Q5: What is the impact of a ground-truth sensor?}}\label{sec:ground-truth-impact} Qualitatively, the ground-truth sensor enables the detection of causality between human action and hidden sensors. Even if all hidden devices were connected to an accessible \wifi network (which is the same system model used by  IoTInspector~\cite{huang2019iot}), one would only be able to detect the presence of a device on the network and not whether it is monitoring a user. To quantitatively demonstrate and evaluate the impact of a ground-truth sensor,  Figure~\ref{fig:cam_traffic} illustrates an example where an IMU enables \name to identify between a hidden sensor monitoring a user and disregard a camera in a separate room. Moreover, one may argue that an application can actively instruct the user to move and establish causality between the period of instruction and the \wifi traffic patterns. First, such an approach relies on a general user motion model to establish causality during these time frames. Second, this approach is not capable of background detection as it would rely on active command and response patterns. In Table~\ref{tab:fpa} case 1, without a ground truth sensor, the false positive rate is 100\%. With a ground truth sensor, this decreases to $6.66$\%.}
\section{Related Work}\label{sec:related}
This section presents the most relevant and related works.

\noindent\textbf{Detecting hidden devices using RF signals.} A popular tool to detect hidden devices is called a bug detector \cite{nbc_2019} -- an RF receiver that can sense if the received power in a frequency range is above a threshold. The problem with such devices is that they can produce false alarms when used near other RF sources such as mobile phones or laptops \cite{spyvsspy, sathyamoorthy2014wireless}. Also, they give no additional information about the type of device or where it is located. After detection, the onus lies completely on the user to physically find the device and verify if it is a surveillance device or not. The host may have a wireless device to monitor the power consumption of his property, but to the bug detector, it would seem similar to an IP camera.

\noindent\textbf{Classifying devices on the network using wireless traffic sniffing.} While services like Princeton IoT Inspector \cite{huang2019iot} collect traffic statistics to identify the types of devices present on the network, they fail to identify if those devices are indeed spying on the user or not. Just ascertaining the presence of a surveillance device is not enough. The device may be present outside the house or it may be monitoring some part of the house which was already disclosed by the home owner. In cases like this, just identifying such a device exists is not enough, we also need to determine two important facets -- is the device spying on the user and is it located in an area of the house that has the potential to violate user privacy. Moreover, tools like this need to have access to the network in order to be effective. If the snooping devices are placed in a hidden network or on a password protected network, the use cases of such a tool are limited. 

Other network traffic analysis tools \cite{solarwinds, schmitt2018enhancing} utilize traffic data to find which devices are consuming high bandwidth. Such techniques can be used to classify audio and video data streams present in the wireless networks. However, with an increase in streaming services \cite{steele_2019, kumar2019all}, it is difficult to distinguish camera video and audio flows with those of streaming services based on just their bandwidth usage.


\noindent\textbf{Detecting cameras on the network using wireless traffic sniffing.} Wampler \etal~\cite{wampler2015information} and others~\cite{nassi2019drones, liu2018detecting} show that information leakage occurs in camera traffic due to how videos are encoded. They observe that changing lighting conditions cause noticeable variations in the network traffic. Though these techniques perform well, their performance degrades when the environment lighting changes naturally. Additionally, while these techniques work well for a camera, they do not generalize to other types of snooping devices, like RF sensors or motion detectors. Finally, in order to be able to change the lighting conditions of a space, the user requires either specialized hardware (like an LED board or a bulb) or access to lighting controls, which is not guaranteed.

Approaches like DewiCam \cite{cheng2018dewicam} exploit the correlation between human motion and camera data flows to determine if the camera is indoors or outdoors.

In \cite{wu2019you}, Wu \etal use their own camera to record a scene while simultaneously sniffing the network traffic. They compare the data rate and pattern of their camera with other devices in the network to look for any similarities. If a similarity exists, there is a high probability that the device is a camera.

\noindent\textbf{Localizing wireless devices using RSSI.} Received Signal Strength Indicator (RSSI) is the estimate of the power received at the receiver from the transmitting device. The power received drops with distance, and so does the RSSI. This property is leveraged to localize devices using RSSI \cite{sun2014wifi, luo2011comparative, xue2017improved, li2018adversarial}. However, due to phenomenon like multipath and shadowing, the accuracy varies from space to space \cite{jondhale2016issues}. The error is very high (several meters). For small rooms, such a result will be meaningless, as the snooping device can be effectively hidden anywhere.
\section{Conclusion}\label{sec:conclusion}
In this paper, we presented \namenospace, a framework to detect, identify, and localize \wifi based sensors monitoring a person in an arbitrary space. \name works by establishing causality between a set of ground truth sensors monitoring a user and the transmitted information of wireless devices on a \wifi network. It then uses this causality to perform trial-based localization. We implement \name on a set of commonly available devices such as a smartphone and a laptop and evaluate our solution on a set of representative clandestine sensors. The framework had a detection rate of $95.2\%$ when the injected multi-modal event was human motion or sound. \name leverages directionality of snooping sensors to reduce the total search area.

\section{Acknowledgements}
The research reported in this paper was sponsored in part by the National Science Foundation (NSF) under award \#CNS-1705135, by the CONIX Research Center, one of six centers in JUMP, a Semiconductor Research Corporation (SRC) program sponsored by DARPA, and by the Army Research Laboratory (ARL) under Cooperative Agreement W911NF-17-2-0196. The views and conclusions contained in this document are those of the authors and should not be interpreted as representing the official policies, either expressed or implied, of the ARL, DARPA, NSF, SRC, or the U.S. Government. The U.S. Government is authorized to reproduce and distribute reprints for Government purposes notwithstanding any copyright notation here on.

\bibliographystyle{IEEEtran}
\bibliography{Sources/references.bib}
\begin{appendices}
\section{Audio-based Localization for Personal Home Assistants}\label{sec:audio_loc}

\begin{figure}[h]
\centering
\includegraphics[width=0.9\linewidth]{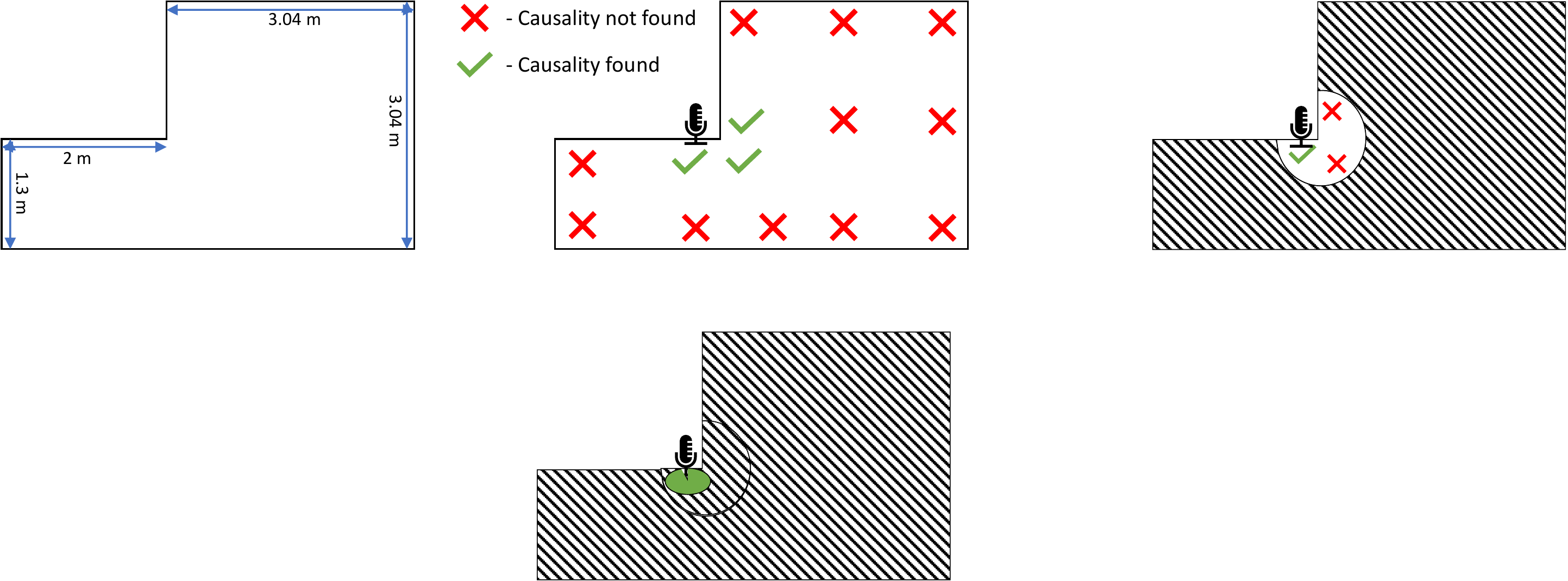}
\caption{Trial-based localization for acoustic sensors.}
\label{fig:audio_loc}
\end{figure}

In this section, we describe the audio localization technique step-by-step. First, we place the source of the sound (smartphone playing a phrase containing the wake word of the device) at different points in the room and see how it affects the device traffic. Then we go around the room while \name repeats that sound continuously and checks them for causality with device traffic as shown in Figure \ref{fig:audio_loc}. Sound played at the points marked as green produces cause-effect relationship with the device traffic. We eliminate the regions where we detect no causality. Next, we reduce the volume by 1 level and repeat our experiment in the left-over space till we are left with a region of desirable size. 

\section{Traffic Variation of a Personal Home Assistant During Drop-In}\label{sec:drop_in}

\begin{figure}[h]
\centering
\includegraphics[width=0.95\linewidth]{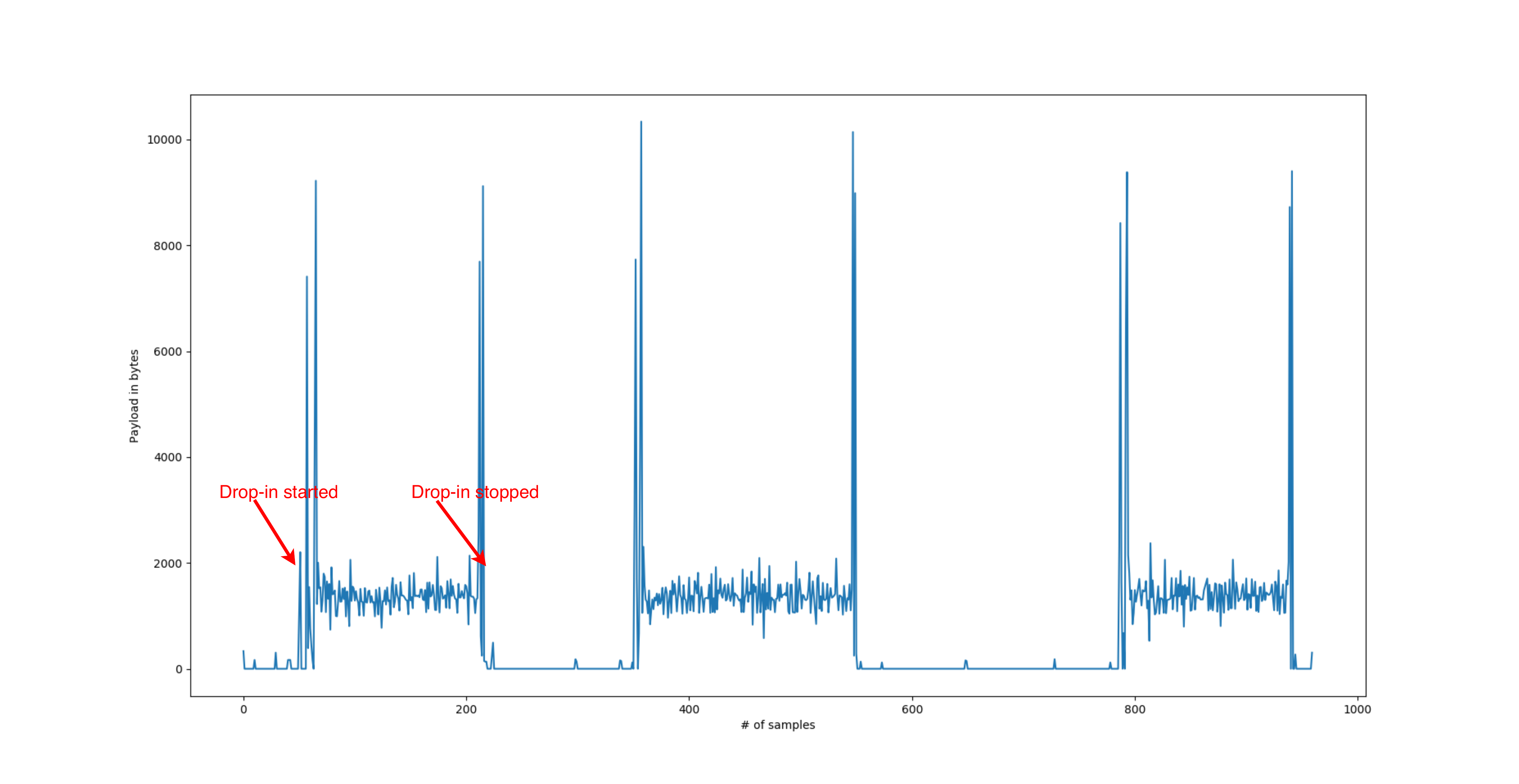}
\caption{\newchange{Traffic variation of Amazon Echo Dot while dropping-in}}
\label{fig:drop_in}
\end{figure}

\newchange{As discussed in the previous sections, Amazon Echo devices allow the user to drop into any of their Alexa devices and remotely listen to the audio in the room that they are placed in. This does not require any authentication on the device side during the drop-in. We perform 3 drop-ins on an Amazon Echo Device and show the traffic variation in Figure \ref{fig:drop_in}. From the traffic variation, it is clear when the drop-ins start and when they end.}

\section{Aggregation of Traffic Statistics}\label{sec:agg}
Each device's traffic is grouped by MAC address, windowed, and processed to compute device traffic volume and variation. \name monitors packet sequence number in the WLAN layer to isolate and remove duplicate or redundant packets. As large images are sent over multiple fixed-length packets, a sufficiently large window size must be used. We chose a 100 ms window to group all packets with the same image within one interval. Cameras require a frame rate higher than $10$ Hz to satisfy the flicker fusion (i.e., persistence of vision) threshold of the human eye \cite{liu2018detecting, green1969sinusoidal}.

For camera encodings, we discard I-frames (through averaging), as they do not encode differences in a scene and require higher bandwidth, thereby adversely affecting the causality analysis.

\end{appendices}
\end{document}